\documentclass[aps,prl,amsmath,amssymb,twocolumn,showpacs,superscriptaddress,final,floatfix]{revtex4-1}
\usepackage{graphicx}	
\usepackage{color}	
\usepackage[normalem]{ulem}
\usepackage{hyperref}
\hypersetup{
  colorlinks,
  citecolor=blue,
  linkcolor=blue,
  urlcolor=blue}

\newcommand{\BFA}{BaFe$_2$As$_2$}

\newcommand{\ie}{{\em i.e.}}

\begin{document}

\title{Momentum Dependence of the Nematic Order Parameter in Iron-based superconductors}

\author{H. Pfau}
\email{hpfau@lbl.gov}
\affiliation{Stanford Institute of Materials and Energy Science, SLAC National Accelerator Laboratory, Menlo Park, California 94025, USA}
\affiliation{Lawrence Berkeley National Laboratory, Berkeley, California 94720, USA}
\author{S. D. Chen}
\affiliation{Geballe Laboratory for Advanced Materials, Department of Applied Physics, Stanford University, Stanford, California 94305, USA}
\author{M. Yi}
\affiliation{Department of Physics, University of California, Berkeley, California 94720, USA}
\affiliation{Department of Physics and Astronomy, Rice University, Houston, Texas 77005, USA}
\author{M. Hashimoto}
\affiliation{Stanford Synchrotron Radiation Lightsource, SLAC National Accelerator Laboratory, Menlo Park, California 94025, USA
}
\author{C. R. Rotundu}
\affiliation{Stanford Institute of Materials and Energy Science, SLAC National Accelerator Laboratory, Menlo Park, California 94025, USA}
\author{J. C. Palmstrom}
\affiliation{Stanford Institute of Materials and Energy Science, SLAC National Accelerator Laboratory, Menlo Park, California 94025, USA}
\affiliation{Geballe Laboratory for Advanced Materials, Department of Applied Physics, Stanford University, Stanford, California 94305, USA}
\author{T. Chen}
\affiliation{Department of Physics and Astronomy, Rice University, Houston, Texas 77005, USA}
\author{P.-C. Dai}
\affiliation{Department of Physics and Astronomy, Rice University, Houston, Texas 77005, USA}
\author{J. Straquadine}
\author{A. Hristov}
\affiliation{Geballe Laboratory for Advanced Materials, Department of Applied Physics, Stanford University, Stanford, California 94305, USA}
\author{R. J. Birgeneau}
\affiliation{Department of Physics, University of California, Berkeley, California 94720, USA}
\author{I. R. Fisher}
\affiliation{Stanford Institute of Materials and Energy Science, SLAC National Accelerator Laboratory, Menlo Park, California 94025, USA}
\affiliation{Geballe Laboratory for Advanced Materials, Department of Applied Physics, Stanford University, Stanford, California 94305, USA}
\author{D. Lu}
\affiliation{Stanford Synchrotron Radiation Lightsource, SLAC National Accelerator Laboratory, Menlo Park, California 94025, USA
}
\author{Z.-X. Shen}
\affiliation{Stanford Institute of Materials and Energy Science, SLAC National Accelerator Laboratory, Menlo Park, California 94025, USA}
\affiliation{Department of Physics, Stanford University, Stanford, California 94305, USA}
\affiliation{Geballe Laboratory for Advanced Materials, Department of Applied Physics, Stanford University, Stanford, California 94305, USA}

\date{\today}


\begin{abstract}

The momentum dependence of the nematic order parameter is an important ingredient in the microscopic description of iron-based high-temperature superconductors. While recent reports on FeSe indicate that the nematic order parameter changes sign between electron and hole bands, detailed knowledge is still missing for other compounds. Combining angle-resolved photoemission spectroscopy (ARPES) with uniaxial strain tuning, we measure the nematic band splitting in both FeSe and BaFe$_2$As$_2$ without interference from either twinning or magnetic order. We find that the nematic order parameter exhibits the same momentum dependence in both compounds with a sign change between the Brillouin center and the corner. This suggests that the same microscopic mechanism drives the nematic order in spite of the very different phase diagrams.

\end{abstract}

\maketitle


Nematicity is increasingly found to be a pervasive feature of strongly correlated systems\cite{ando_2002,kasahara_2006,achkar_2016,ronning_2017}. It is therefore important to understand its microscopic mechanism in order to determine its relation to other quantum phenomena, in particular, unconventional superconductivity.
The phase diagram of the majority of iron-based superconductors (FeSCs) contains a nematic phase  \cite{paglione_2010,johnston_2010,kuo_2016}. It is often accompanied by a spin-density wave (SDW) phase and spin fluctuations were proposed as its driving force \cite{fernandes_2014}. In contrast, the discovery of nematicity without long-range magnetism in FeSe promoted orbital fluctuations as the driving mechanism \cite{beak_2014,boehmer_2014}. It is currently being debated whether there is a common microscopic mechanism of nematic order in FeSC. 

The nematic phase transition involves (1) a change from tetragonal to orthorhombic crystal structure, (2) an in-plane anisotropy of the spin susceptibility, and (3) an anisotropic occupation of $d_{xz}$ and $d_{yz}$ orbitals with an energy shift of the corresponding bands in opposite direction \cite{fernandes_2014,paglione_2010,johnston_2010,yi_2017_npj}. In a Ginzburg-Landau description of the free energy, the nematic phase transition is characterized by an order parameter $\phi_0$, which becomes nonzero inside the nematic state. Importantly, one can define a momentum-dependent order parameter $\phi_\mathrm{nem}(k) = \phi_0 f(k)$ that contains information about the microscopic mechanisms behind nematic order similarly to the momentum-dependent superconducting order parameter. Experimentally, the nematic band splitting $\Delta E_\mathrm{nem}$, which defines the anisotropy of the dispersion between $k_x$ and $k_y$, gives access to $\phi_\mathrm{nem}(k)$. Its comparison between FeSe and FeSCs with magnetic order will give important insights into the question of a common driving force of nematicity.

FeSe undergoes a nematic phase transition at 90\,K \cite{mcqueen_2009}. Only a few studies have reported on the momentum dependence of the order parameter: ARPES on thin films indicates a strong momentum dependence of the nematic band splitting \cite{zhang_2016}, while the Fermi surface distortion observed in detwinned crystals reveals a sign change between hole and electron bands \cite{suzuki_2015}.
 
In contrast to FeSe, it is nontrivial to disentangle the effects of nematicity and SDW order in most other FeSC. Both orders appear almost simultaneously below $T_\mathrm{nem} = 137$\,K in the prototype \BFA \cite{kim_2011_2}. As a result, the low-temperature electronic structure is affected by complex effects from both nematicity and magnetic order \cite{hsieh_2008,yang_2009,yi_2009_prb,yi_2011_pnas,kim_2011,wang_2013,liu_2009_prb,kondo_2010,jensen_2011,pfau_2019_prb,richard_2010,shimojima_2010} and the detailed momentum profile of the nematic order parameter remains unclear.

Here we report on the in-plane momentum dependence of the nematic band splitting in FeSe and \BFA~determined by angle-resolved photoemission spectroscopy (ARPES). We study FeSe at $T<T_\mathrm{nem}$ and compare the dispersion along two orthogonal directions in a detwinned single crystal \cite{yi_2011_pnas}. The interference from SDW order in \BFA~requires a different approach. We apply an {\it{in-situ}} tunable uniaxial pressure along the Fe-Fe bond direction. The antisymmetric component of the resulting strain couples to the electronic nematic order parameter \cite{kuo_2014}. We demonstrate, that such a strain induces a nematic band splitting at temperatures above $T_\mathrm{nem}$, while the system remains in the paramagnetic phase. This establishes strain as a continuous {\it{in-situ}} tuning parameter for photoemission spectroscopy in FeSC. We find a strong momentum dependence of $\Delta E_\mathrm{nem}$ and correspondingly $\phi_\mathrm{nem}$ including a sign change between the center and the corner of the Brillouin zone (BZ). The functional form is the same for FeSe and \BFA, in spite of the dramatic differences in the behaviors of the magnetic order.  Our results, therefore, suggest that the same microscopic mechanism governs the nematic phase in FeSC with and without magnetic order.


High-quality single crystals of \BFA~and FeSe were grown using self-flux and chemical vapor transport methods, respectively \cite{chu_2009,wang_2009,rotundu_2010}. ARPES measurements were performed at the SSRL beamline 5-2 with an energy and angular resolution of 12\,meV and 0.1$^\circ$. The samples were cleaved {\it{in-situ}} with a base pressure below $5\times 10^{-11}$\,torr. The chosen photon energies of 37\,eV for FeSe and of 47\,eV for \BFA~probe a $k_z$ close to the BZ center \cite{watson_2015,brouet_2009}. Different orbital contributions were highlighted using linear horizontal (LH) and linear vertical (LV) light polarization \cite{zhang_2011_prb,brouet_2012,yi_2011_pnas}.

\begin{figure}
\begin{center}
\includegraphics[width=\columnwidth]{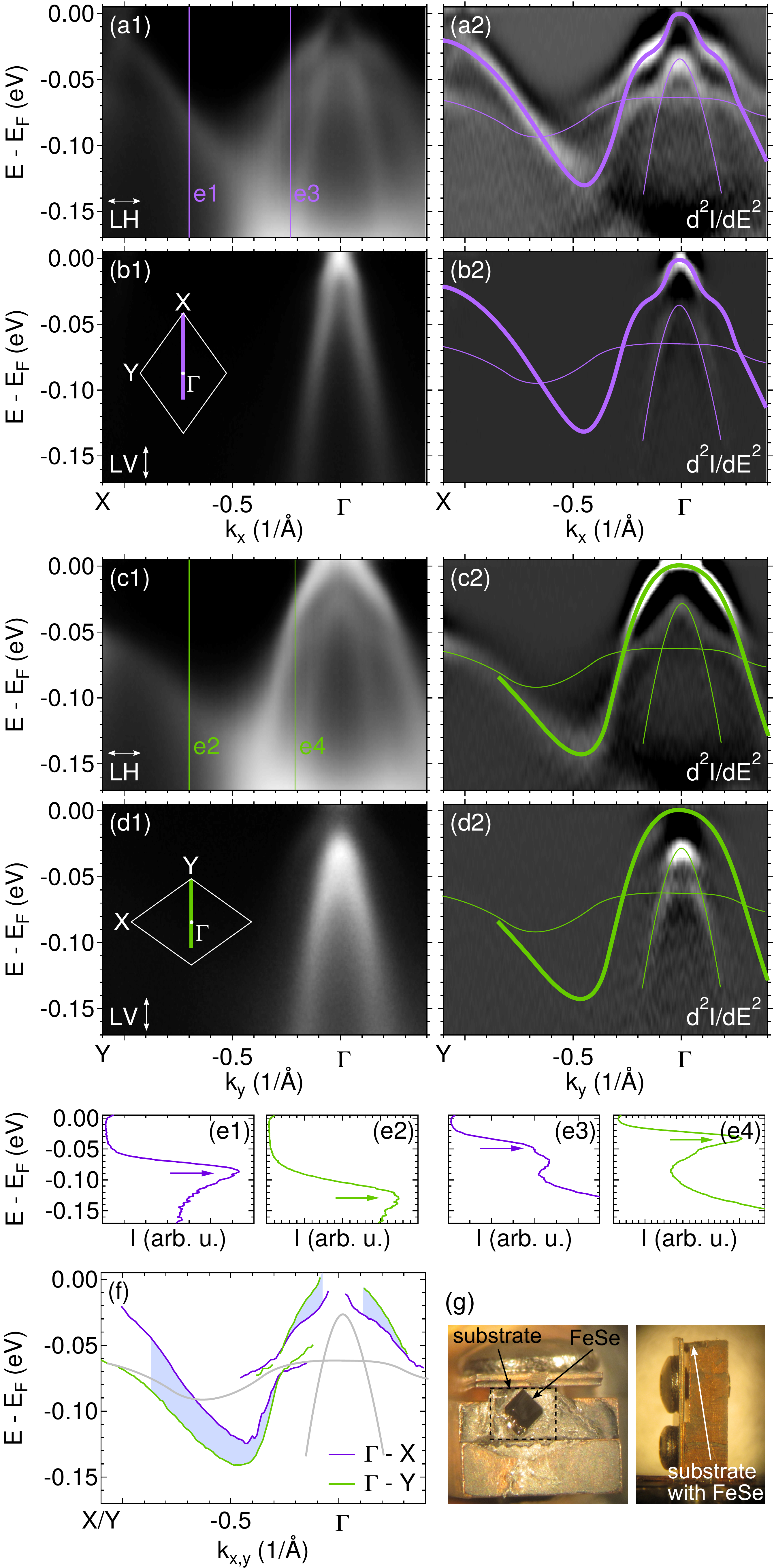}
\caption{
Detwinned FeSe at 15\,K. (a,b) Spectra along the $\Gamma$--X direction, \ie~perpendicular to the applied pressure, for LH and LV polarization. (a1,b1) show the ARPES spectra divided by the Fermi function. (a2,b2) depict their second energy derivative. Lines mark the dispersions as guides to the eye. (c,d) same as (a,b) for the $\Gamma$--Y direction, \ie~parallel to the applied pressure. (e) Selected EDCs for momenta marked in (a1,c1). The arrows highlight the binding energy of the $d_{yz}$ (e1,e3) and $d_{xz}$ (e2,e4) hole band. (f) Band dispersion extracted from minima in second derivative of EDCs for $\Gamma$--X and $\Gamma$--Y direction together with guides for the dispersion of the other two hole bands (gray lines). Shaded area highlights the nematic band splitting $\Delta E_\mathrm{nem}$. (g) Photographs of the mechanical detwinnning clamp.
}
\label{Fig:FeSe}
\end{center}
\end{figure}

\begin{figure}
\begin{center}
\includegraphics[width=\columnwidth]{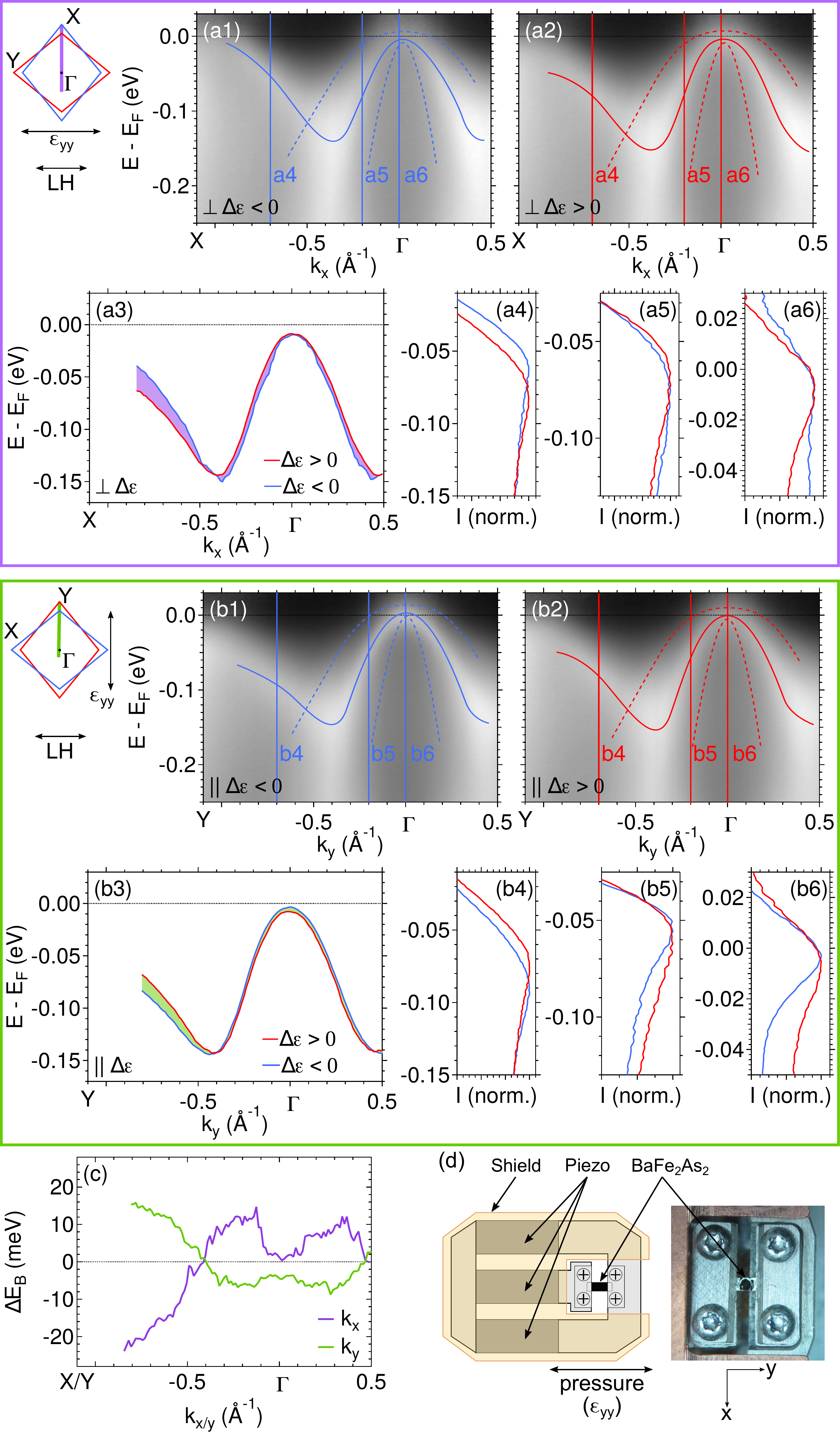}
\caption{
Strained \BFA~at 160\,K. (a) Data along a momentum perpendicular to strain direction. ARPES spectra divided by the Fermi function under (a1) compressive strain and (a2) tensile strain. Lines indicating band dispersions are guides to the eye. (a3) Dispersion of the center hole band (solid line in (a1,a2)) for tensile (red) and compressive (blue) strain extracted from maxima in EDCs. (a4-a6) EDCs at momenta marked in (a1,a2). Intensity $I$ normalized to maximum value. (b) Same as (a) for momentum direction parallel to the applied strain. (c) Difference of the band dispersion $\Delta E_\mathrm{B}$ between tensile and compressive strain for momentum directions parallel and perpendicular to the applied strain. (d) Sketch of the strain apparatus and photograph of the mounted sample.
}
\label{Fig:Ba122}
\end{center}
\end{figure}

We studied FeSe at $15\,\mathrm{K} < T_\mathrm{nem}$ and used a mechanical clamp as shown in Fig.~\ref{Fig:FeSe}(g) for detwinning. It applied pressure to a substrate made from \BFA, onto which we glued the FeSe crystal. Previous neutron scattering experiments revealed that FeSe can be completely detwinned using this method \cite{chen_2019}.
\BFA~was studied at $160\,\mathrm{K} > T_\mathrm{nem}$. A strain device with piezoelectric stacks as shown in Fig.~\ref{Fig:Ba122}(d) was used to apply an {\it{in-situ}} tunable uniaxial pressure \cite{hicks_2014}. We confirmed, that a metallic shielding prevents the high voltage of the piezoelectric stacks to alter the ARPES measurement. We compared spectra taken with compressive and tensile pressure that correspond to +(-)90\,V applied to the center(outer) piezoelectric stacks and vice versa. We used the same pressure for both studied momentum directions. A strain gauge was used to estimate the strain between both settings to be $\Delta l/l \approx 0.16$\%. No signatures of strain-induced magnetic order were observed at 160\,K.

Recently, strain-dependent ARPES studies on different materials were reported employing mechanical mechanisms that either bend \cite{ricco_2018} or stretch \cite{flototto_2018} a substrate onto which a sample is glued. While these mechanical devices do not require electric shielding, a piezoelectric device allows us to continuously and reversibly tune compressive and tensile strain and to measure the applied strain at the same time.


\begin{figure}
\begin{center}
\includegraphics[width=\columnwidth]{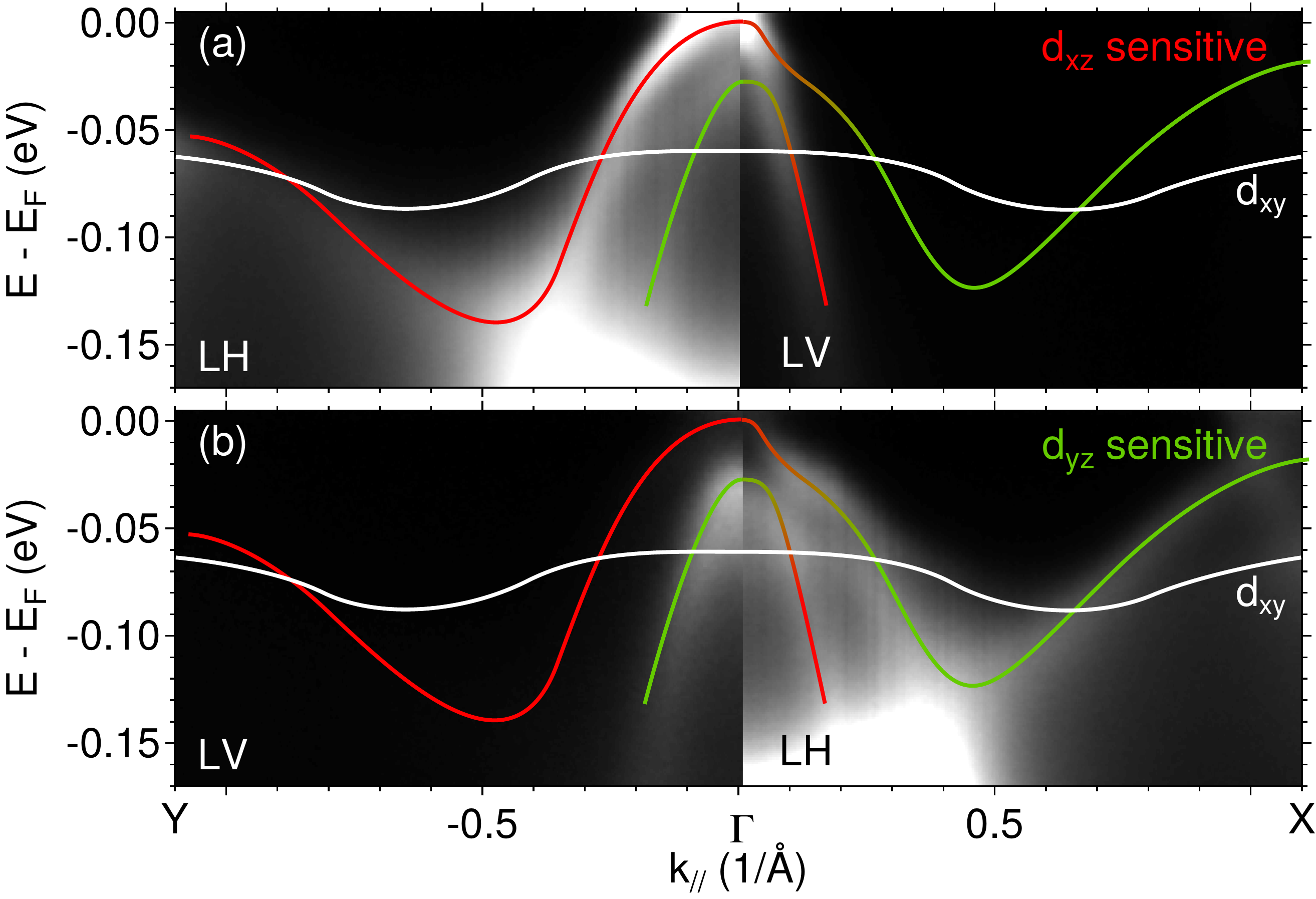}
\caption{Orbital redistribution. ARPES spectra of FeSe at 15\,K along Y--$\Gamma$--X. (a) highlights contributions from $d_{xz}$ orbitals (left: LH, right: LV polarization) and (b) those of $d_{yz}$ orbitals (left: LV, right: LH polarization). Lines are guides to the eye colored corresponding to orbital character.}
\label{Fig:orbital}
\end{center}
\end{figure}

\begin{figure}
\begin{center}
\includegraphics[width=\columnwidth]{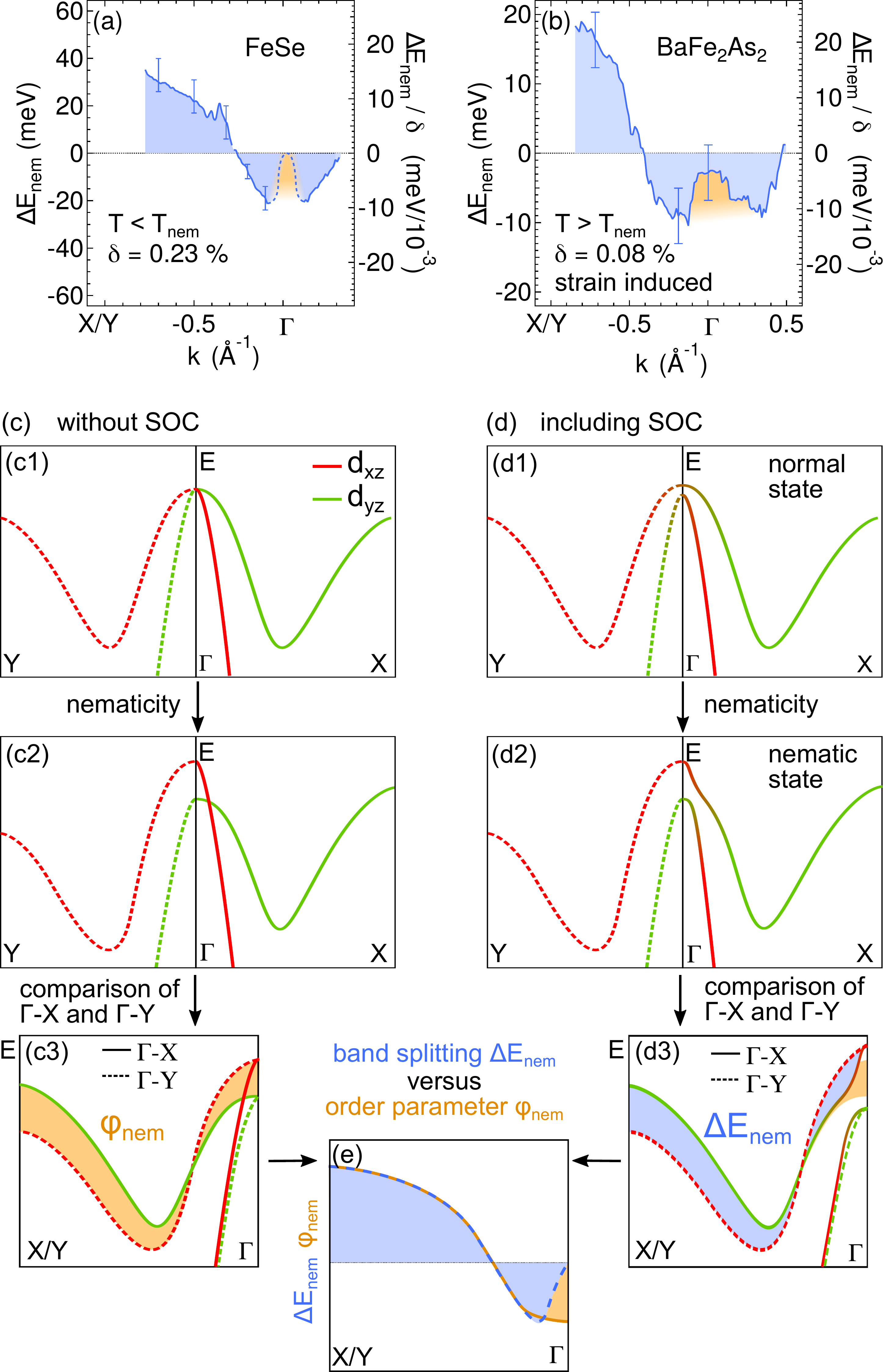}
\caption{
Nematic band splitting $\Delta E_\mathrm{nem}$ and nematic order parameter $\phi_\mathrm{nem}$. (a,b) $\Delta E_\mathrm{nem}$ of FeSe and \BFA~as function of momentum. We normalized $\Delta E_\mathrm{nem}$ by the orthorhombic distortion $\delta$ of each sample for the right axes. Representative error bars are included. Dashed line in (a) is a guide to the eye. (c,d) Sketches of the changes in the band structure due to nematicity excluding (c) and including (d) effects of SOC. Only bands with $d_{xz}$ and $d_{yz}$ orbital characters are shown for simplicity. We define the nematic order parameter $\phi_\mathrm{nem}$ in (c3).  We sketch the difference to the measured $\Delta E_\mathrm{nem}$ determined from (d3) in (e).}
\label{Fig:nem_split}
\end{center}
\end{figure}

Figure \ref{Fig:FeSe} summarizes our ARPES results on detwinned FeSe. We identify three hole bands centered at $\Gamma$, which we highlight with lines on top of the second derivative spectra Fig.~\ref{Fig:FeSe}(a2-d2). Signatures of hybridization become visible at points where they cross. In the following, we focus our analysis on the middle hole band marked with thick lines. It has $d_{yz}$ character along the $\Gamma$--X direction and $d_{xz}$ character along $\Gamma$--Y. They are intense in LH polarization and suppressed in LV polarization, consistent with the orbital character assignment. We extract their dispersion from minima in the second derivative of the energy distribution curves (EDCs) and plot them in Fig.~\ref{Fig:FeSe}(f). We find a binding energy difference $\Delta E_\mathrm{nem}$ along the two orthogonal momentum directions. This difference is a signature of the nematic order. We extract the momentum dependence of $\Delta E_\mathrm{nem}$ and plot it in Fig.~\ref{Fig:nem_split}(a). We only include momenta highlighted by the shaded area in Fig.~\ref{Fig:FeSe}(f). The band assignment in this momentum region agrees with existing literature \cite{watson_2015, suzuki_2015, fanfarillo_2016,yi_2019_arxiv}. We disregard the region beyond $k=-0.8\,\mathrm{\mathring{A}^{-1}}$, at which the $d_{xz}$ band touches the $d_{xy}$ band along $\Gamma$--Y. Beyond this momentum close to the BZ corner, the band assignment and the resulting nematic band splitting is currently debated in literature due to this band crossing \cite{yi_2019_arxiv,fedorov_2016,watson_2016,zhang_2016}. For small momenta close to $\Gamma$, the dispersion could not be determined reliably when the band is too close to or above $E_\mathrm{F}$.  Our spectra in Fig.~\ref{Fig:FeSe}(a-d) confirm previous results on detwinned FeSe \cite{suzuki_2015}. While Ref.~\onlinecite{suzuki_2015} focuses on the Fermi surface distortions, we extended this work by extracting the full momentum dependence of the nematic splitting.

The nematic band splitting leads to a redistribution of orbital character at $E_\mathrm{F}$. We show the spectra of FeSe in the ordered state arranged according to the orbital sensitivity of different light polarization in Fig.~\ref{Fig:orbital}. We observe an increase of $d_{xz}$ orbital character at the Fermi level around $\Gamma$ while the $d_{yz}$ spectral weight is pushed below $E_\mathrm{F}$. Figure \ref{Fig:orbital} also reveals a hybridization signature along  $\Gamma$--X close to $\Gamma$. It is a consequence of spin-orbit coupling (SOC) and underlines the importance of this interaction for the understanding of the electronic structure in FeSC \cite{fernandes_2014_prb,borisenko_2015,day_2018,watson_2015}. We will describe its effect in detail below (Fig.~\ref{Fig:nem_split}).


Our experimental observations on strained \BFA~are presented in Fig.~\ref{Fig:Ba122}. In general, uniaxial pressure applied along one of the in-plane Fe-Fe bond directions $x$ or $y$ results in a symmetric strain $(\epsilon_{yy}+\epsilon_{xx})/2$ and an antisymmetric strain $(\epsilon_{yy}-\epsilon_{xx})/2$ (note that here the $x$ and $y$ axes are at $45^\circ$ to the tetragonal [100] and [010] directions) \cite{palmstrom_2017,ikeda_2018}. The antisymmetric strain breaks the same $B_{2g}$ symmetry as the nematic order and, hence, induces a nonzero value of the nematic order parameter at all temperatures. The associated nematic band splitting is given by the antisymmetric term $\Delta E_\mathrm{nem}(k) = [\Delta E_{\mathrm{B}}(k_y) - \Delta E_{\mathrm{B}}(k_x)]/2$, while the symmetric $A_{1g}$ response is given by $\Delta E_\mathrm{sym}(k) = [\Delta E_{\mathrm{B}}(k_y) + \Delta E_{\mathrm{B}}(k_x)]/2$. Here, $\Delta E_\mathrm{B}$ refers to the strain-induced change in binding energy.

Following this idea, we applied uniaxial pressure to one main in-plane axis, which we call $y$ without loss of generality (see Fig.~\ref{Fig:Ba122}(d)). ARPES measurements were performed along $k_x$ and $k_y$, \ie~perpendicular (Fig.~\ref{Fig:Ba122}(a)) and parallel (Fig.~\ref{Fig:Ba122}(b)) to the direction of the applied pressure. For each momentum direction, we took a spectrum for a compressed ($ \epsilon_{yy} <0$) and a tensioned ($\epsilon_{yy} >0$) state of the sample to extract the strain-induced $\Delta E_\mathrm{B}$. We will focus again on the middle hole band (solid line in Fig.~\ref{Fig:Ba122}(a1,a2,b1,b2)) and present spectra taken with LH polarization. The dispersion of the hole band is extracted from the maxima of the EDCs and plotted in Fig.~\ref{Fig:Ba122}(a3,b3). We can indeed observe a strain-induced change in the dispersion between the spectra taken with positive and negative pressure and extract the difference $\Delta E_\mathrm{B}$, which is plotted in Fig.~\ref{Fig:Ba122}(c). $\Delta E_\mathrm{B}$ has a very similar functional form for both momentum directions but with opposite sign. The size of the symmetric signal $E_\mathrm{sym}$ is therefore very small and below 5\,meV for all momenta. The nematic band splitting $ \Delta E_\mathrm{nem}$, \ie~the antisymmetric component, is shown in Fig.~\ref{Fig:nem_split}(b). 

The spectra in Fig.~\ref{Fig:Ba122} indicate a strain offset, \ie~0\,V on the piezoelectric stacks does not correspond to $\epsilon_{yy} = 0$, likely due to different thermal expansion coefficients of the different materials. The offset does not affect our analysis of $\Delta E_\mathrm{nem}$, because it only considers the relative strain and binding energy differences.


Figure \ref{Fig:nem_split}(a,b) compares the results of the nematic band splitting $\Delta E_\mathrm{nem}$ for FeSe and \BFA. $\Delta E_\mathrm{nem}$ has the same momentum dependence with a sign change between $\Gamma$ and the BZ corner. It has a value close to zero at $\Gamma$. For FeSe, no values could be obtained close to $\Gamma$ as detailed earlier, but we expect $\Delta E_\mathrm{nem}=0$ at $\Gamma$ from the trend in the dispersions (Fig.~\ref{Fig:FeSe}). 

Figures \ref{Fig:nem_split}(c,d) sketch the changes of the band structure due to nematic order based on the observed $\Delta E_\mathrm{nem}$. We illustrate the cases without (c) and with (d) SOC. From the case without SOC (Fig.~\ref{Fig:nem_split}(c3)), one can define a momentum-dependent nematic order parameter $\phi_\mathrm{nem}(k_x,k_y)$ as the difference in binding energy between two orthogonal momentum directions. This definition naturally captures the $d$-wave distortion of the Fermi surface that originates from the sign change of the order parameter between $k_x$ and $k_y$. For the high-symmetry direction shown in Fig.~\ref{Fig:nem_split}(c3), our definition implies that $\phi_\mathrm{nem}(k)$ is the difference between the binding energies along $k_y$ and along $k_x$.

The comparison to Fig.~\ref{Fig:nem_split}(d) underlines the large effect of SOC on the band structure and the measured $\Delta E_\mathrm{nem}(k)$. Importantly, the experimentally determined quantity $\Delta E_\mathrm{nem}(k)$  equals $\phi_\mathrm{nem}(k)$  only away from $\Gamma$ where SOC does not affect the dispersion, see Fig.~\ref{Fig:nem_split}(e). In particular, the crossing of the $d_{xz}$ and $d_{yz}$ band close to $\Gamma$ along $\Gamma$--X leads to a hybridization gap and $\Delta E_\mathrm{nem}(\Gamma) = 0$ while $\phi_\mathrm{nem}(\Gamma)$ stays nonzero. However, from the comparison of $\phi_\mathrm{nem}$ and $\Delta E_\mathrm{nem}$ in Fig.~\ref{Fig:nem_split}(e) and from the functional form of $\Delta E_\mathrm{nem}$ in Fig.~\ref{Fig:nem_split}(a,b), we can conclude that the nematic order parameter $\phi_\mathrm{nem}(k_x,k_y)$ has an additional sign change between the BZ center and the BZ corner close to $|k|=0.3 \mathring{A}^{-1}$. At this wave vector, $\Delta E_\mathrm{nem}$ is not altered by the effect of SOC that is described above. $|\phi_\mathrm{nem}|$ is approximately twice as large at the BZ corner compared to the BZ center. To determine the exact value at $\Gamma$, SOC has to be taken into account. One expects a binding energy difference between the inner and middle hole band at $\Gamma$ of $ (\lambda^2 + \phi_{\mathrm{nem}}^2(\Gamma))^{1/2}$ \cite{fernandes_2014_prb}, which is approximately 30\,meV for FeSe determined from Fig.~\ref{Fig:FeSe}. Together with a SOC of $\lambda = 25$\,meV \cite{borisenko_2015} we obtain $\phi_{\mathrm{nem}}(\Gamma) \approx 17$\,meV in agreement with the results shown in Fig.~\ref{Fig:nem_split}(a). 

We point out that the center hole band loses $d_{xz}$ or $d_{yz}$ character away from high-symmetry points of the BZ. However, they remain the majority orbital contribution \cite{graser_2010}. Such orbital admixture can lead to modulations of $\Delta E_\mathrm{nem}(k)$. However, the sign change of $\phi_\mathrm{nem}(k)$ between $\Gamma$ and the BZ corner is robust against these considerations because the bands have pure orbital character at the high-symmetry points and the band shift is therefore proportional to $\phi_\mathrm{nem}(k)$.

Given the different orthorhombic distortions $\delta$ in FeSe and \BFA, we also compare the nematic susceptibility $\Delta E_\mathrm{nem}/\delta$. From the lattice constants of FeSe inside the nematic state \cite{horigane_2009} we determine $\delta = (a-b)/(a+b) = 0.23$\,\%. We estimate $\delta = \Delta l/2l = 0.08$\% for \BFA~from the strain measurement, which is smaller than the orthorhombic lattice distortion of 0.39\% inside the nematic phase \cite{huang_2008}. Therefore, the magnitude of $\Delta E_\mathrm{nem}$ in \BFA~will be larger inside the nematic phase compared to Fig.~\ref{Fig:nem_split}(b). The nematic susceptibility has the same order of magnitude in both compounds, demonstrating a similar strength of the nematic order. Small differences are expected: First,  $\Delta E_\mathrm{nem} / \delta$ is temperature dependent \cite{chu_2012,boehmer_2014,massat_2016} and we compare measurements at different temperatures $T/T_\mathrm{nem}$. Second, in different optimally doped FeSC, the nematic susceptibility has slightly different values \cite{kuo_2016}.

Our experimental result demonstrates, that the nematic order parameter has the same momentum dependence in both FeSe and \BFA~and could even be  universal among the FeSCs. It suggests that  the same microscopic mechanism drives  nematicity despite the absence of magnetism in FeSe. A pure on-site ferro-orbital order can be excluded for both systems as it does not support a sign change in $\Delta E_\mathrm{nem}$. A number of other models were proposed including bond-orbital order \cite{su_2015,mukherjee_2015,li_2017,onari_2016}, a Pomeranchuk instability \cite{chubukov_2016}, orbital-selective spin fluctuations \cite{fanfarillo_2016}, frustrated magnetism \cite{wang_2015}, and spin-driven Ising-nematic order \cite{fernandes_2010,yu_2015}. 
Our result puts strong constraints on the theoretical description of the nematic order in FeSC.


In summary, we determined the momentum dependence of the nematic order parameter in FeSe and in \BFA~using ARPES. To this end, we studied the nematic band splitting in detwinned FeSe inside the ordered phase. \BFA~was studied above the magnetic ordering temperature and we induced a nematic band splitting in a controlled way by the application of uniaxial pressure. Despite the very different magnetic properties of both materials, the nematic order parameter exhibits the same momentum dependence with a sign change between the BZ center and the BZ corner. It will be very interesting to perform similar studies on other nematic FeSC to test if this is a universal behavior of the nematic order parameter.

\begin{acknowledgments}
We are very grateful for valuable discussions with R. Fernendes, A. Kemper, Q. Si, Y. Zhang and C. Watson. H.P. acknowledges support from the Alexander von Humboldt Foundation and from the German Science Foundation (DFG) under reference PF 947/1-1. J.C.P. is supported by a Gabilan Stanford Graduate Fellowship and a NSF Graduate Research Fellowship (grant DGE-114747). J.S. acknowledges support from an ABB Stanford Graduate Fellowship. This work was supported by the Department of Energy, Office of Basic Energy Sciences, under Contract No. DE-AC02-76SF00515. The FeSe single crystal growth and characterization work at Rice was supported by the U.S. DOE, BES DE-SC0012311 and in part by the Robert A. Welch Foundation Grant No. C-1839 (P.-C.D.). Use of the Stanford Synchrotron Radiation Lightsource, SLAC National Accelerator Laboratory, is supported by the U.S. Department of Energy, Office of Science, Office of Basic Energy Sciences under Contract No. DE-AC02-76SF00515. Work at Lawrence Berkeley National Laboratory was funded by the U.S. Department of Energy, Office of Science, Office of Basic Energy Sciences, Materials Sciences and Engineering Division under Contract No. DE-AC02-05-CH11231 within the Quantum Materials Program (KC2202).
\end{acknowledgments}


\bibliography{nematic_splitting_manuscript}

\end{document}